\documentclass[aps,prl,twocolumn,showpacs,preprintnumbers]{revtex4}
\usepackage{amssymb}
\usepackage{multirow}
\usepackage[colorlinks,linkcolor=blue,citecolor=blue,dvipdfm]{hyperref}
\usepackage{graphicx}
\usepackage{dcolumn}
\usepackage{bm}
\usepackage{color,soul}

\begin{document}

\title{Resonance, Fermi surface topology, and Superconductivity in Cuprates}
\date{\today}

\begin{abstract}
The resonance, a collective boson mode, was usually thought to be a possible glue of superconductivity. We argue that it is rather a natural product of the \emph{d}-wave pairing and the Fermi surface topology. A universal scaling $E_{res}/2\Delta ^{H}_{S}\sim 1.0$ ($\Delta_{S}^{H}$ the magnitude of superconducting gap at hot spot) is proposed for cuprates , irrespective of the hole-/electron-doping, low-/high-energy
resonance, monotonic/nonmonotonic \emph{d}-wave paring, and the parameters selected. We reveal that there may exist two resonance peaks in the electron-doped cuprates. The low- and high- energy resonance, originated from the contributions of the different
intra-band component, is intimately associated with the Fermi surface topology. By analyzing the data of inelastic neutron scattering, we conclude the nonmonotonic \emph{d}-wave superconducting pairing symmetry in the electron-doped cuprates, which is still an open question.

\end{abstract}

\author{Y. Zhou$^1$, H. Y. Zhang$^1$, H. Q. Lin$^2$, and C. D. Gong$^{3,1}$ }
\affiliation{$^1$National Laboratory of Solid State Microstructure, Department of
Physics, Nanjing University, Nanjing 210093, China \\
$^{2}$Beijing Computational Science Research Center, Beijing 100084, China\\
$^3$Center for Statistical and Theoretical Condensed Matter Physics,
Zhejiang Normal University, Jinhua 321004, China }
\maketitle

\date{\today }

High-temperature superconductivity, arising from the charge carrier doping
into their parent compounds, is one of the most challenging topics in condense
matter physics\cite{Armitage-RMP}. The fundamental issue is what the glue of
pairing is in cuprates. The superconductivity is generally believed to
associate with some collective boson modes. For example, due to the
proximity of antiferromagnetism and superconductivity, the spin fluctuations
is often proposed to be the glue of pairing in the cuprates\cite{Jin-Nature11}. Understanding the
nature of the spin fluctuation and its relation to the superconductivity are
essential for the mechanism of High-$T_{c}$ cuprates. The inelastic neutron
scattering(INS) provides a direct way to investigate the spin dynamics.

So far, the extensive INS experiments had been performed on cuprates\cite%
{Armitage-RMP,FujitaRev}. The significant differences of magnetic excitations had been found,
i.e., the 'hour-glass' and low-energy commensurate dispersion of magnetic excitation in hole- and electron-doped cuprates. However, a universal feature in the
superconducting (SC) state is proposed. The resonance\cite%
{Zhao-NPHY11,Fujita-PRL08,Yamada-PRL03,Zhao-PRL07,Wilson-Nature06,Wilson-PNAS07,Yu-NPHY09,Yu-PRB10}%
, a spin-triplet collective mode at $\mathbf{Q}=(\pi ,\pi )$, is discovered
in both type of cuprates. The resonance energy $E_{res}$ is about $9.5meV$,
and $11meV$, in the optimal doping $NCCO$\cite{Zhao-PRL07}, and $PLCCO$\cite%
{Wilson-Nature06}, respectively. This value is much enhanced in the
hole-doped cuprates, e.g., $18meV$ in optimal doped $LSCO$, $47meV$ in $%
Tl2212$, $56meV$ in $Hg1201$, etc (Detailed experimental data, see Ref.\cite%
{Yu-NPHY09}). It is proposed that the ratio of $E_{res}/k_{B}T_{c}$ is fixed
with the value about $5.8$\cite{Wilson-Nature06,Zhao-PRL07}. However, Yu \emph{%
et al.} argued that the universal scaling is $E_{res}/2\Delta_{S}^{AN}\sim0.64$ ($\Delta
_{S}^{AN}$ the SC gap at antinode) instead\cite{Yu-NPHY09}, which is also supported by the scanning
tunnelling microscopy measuremtns\cite{Niestemski-Nature07}. Furthermore,
the single resonance may be constituted by two separated sub-peaks in
optimal doped NCCO\cite{Yu-PRB10}, which seems to conflict with the above mentioned universal ratio. Nevertheless, the resonance is intimately related to the superconductivity. In this sense, the resonance is the essence of the superconductivity and possibly the candidate of the glue of superconductivity.

Based on the kinetic energy driven SC mechanism, a dome shaped doping
dependent resonance energy is proposed in electron-doped cuprates. However,
the intensity at given energy in the SC state is almost three orders of
magnitude lager than that of the normal state\cite{Feng}. Ismer \emph{et al.} showed that the
resonance can be regarded as an overdamped collective mode located near the
particle-hole continuum\cite{Ismer-PRL07}. Their results indicated that the
resonance energy is sensitive to the parameters selected. Furthermore, those
theories based on the single band description\cite{LJX-PRB03,Kruger-PRB07}
could not be able to take account of the properties of spin dynamics, for example, the
commensurate magnetic excitation in electron-doped cuprates, as we argued in
the previous paper\cite{HY}. To our knowledge, the possible linear scaling
of resonance and its intrinsic relation to superconductivity have not
been well established.

In this paper, the resonance is studied in details in the cuprates. The
resonance energy linearly depends on the SC gap. A low-energy resonance
emerges when the the hole-pocket develops in the electron-doped cuprates. The main
features of the magnetic excitations are qualitatively consistent with the
INS measurements. The resonance, coming from respective intra-band
component, is dominated by the scattering between two hot spots. We argue
that the resonance is a natural product of the \emph{d}-wave pairing
symmetry and Fermi surface topology rather than the glue of
superconductivity. A universal scaling $E_{res}/2\Delta ^{H}_{S}\sim 1.0$ with $\Delta ^{H}_{S}$ the magnitude of superconducting gap at the hot spot, which is insensitive to the selected details,
is proposed instead of the experimentally suggested $E_{res}/2\Delta_{S}^{AN}\sim
0.64$. We conclude that the SC
pairing symmetry in electron-doped cuprates is the nonmonotonic \emph{d}-wave. This manifests that the INS measurement can not only be used to judge
the pairing symmetry, but also provides the details.

We start from the generic model with non-zero $\mathbf{Q}$-commensurate
density wave, which assumes a $\mathbf{Q}=\left( \pi ,\pi \right) $ vector
with an energy gap $\Delta _{k}^{N}$ that either is isotropic or has \emph{d}%
-wave symmetry\cite{Norman-PRB07}. The Green's function is%
\begin{equation}
G_{k}^{-1}=\omega -\epsilon _{k}+i\Gamma -\frac{\left( \Delta
_{k}^{N}\right) ^{2}}{\omega -\epsilon _{k+Q}+i\Gamma }
\end{equation}%
with $\epsilon _{k}$ the dressed tight-binding (TB) dispersion and $\Gamma $ the
elastic broadening factor. This nonzero $Q$ scenario, originated from the
spin density wave\cite{Das-PRB85}, charge density wave\cite{Greco-PRL09},
d-density-wave (DDW)\cite{Chakravarty-PRB01}, or spin-spin correlations (SSC)%
\cite{ZLG-JPSJ13} etc., well models the Fermi surface topology in the
underdoped cuprates. In the superconducting state, we introduce a
phenomenological pairing term $-\sum_{k}\Delta ^{k}_{S}\left( c_{k\uparrow
}c_{-k\downarrow }+h.c.\right) $, where $\Delta ^{k}_{S}=\frac{1}{2}\Delta
_{S}(\cos k_{x}-\cos k_{y})$ is the monotonic \emph{d}-wave SC order
parameter usually. However, a nonmonotonic \emph{d}-wave with third harmonic
term\cite{Matsui,Blumberg} $\Delta ^{k}_{S}=\frac{1}{2}[\Delta _{S}(\cos
k_{x}-\cos k_{y})-\Delta _{S}^{\prime }(\cos 3k_{x}-\cos 3k_{y})]$ is
further discussed in the electron-doped case with $\Delta _{S}^{\prime }\sim
\Delta _{S}/2.41$\cite{Kruger-PRB07}. The quasiparticle dispersion is $%
E_{k}^{\eta }=\sqrt{(\xi _{k}^{\eta })^{2}+\left( \Delta ^{k}_{S}\right) ^{2}%
}$ with $\xi _{k}^{\eta }=(\frac{\epsilon _{k}+\epsilon _{k+Q}}{2})+\eta
\sqrt{\left( \frac{\epsilon _{k}-\epsilon _{k+Q}}{2}\right) ^{2}+\left(
\Delta _{k}^{N}\right) ^{2}}$ ($\eta =1$, and $-1$ for upper, and lower
band).\bigskip

We take the phenomenological SDW description for representation, which
reproduces the hourglass-shape\cite{Das-PRB85}, and low-energy commensurate magnetic excitations\cite{HY} in hole-doped, and electron doped cuprates, respectively. The other possibility will be also
discussed. The model Hamiltonian yields to%
\begin{equation}
H=\sum_{k\sigma }\left( \epsilon _{k}-\mu \right) c_{k\sigma }^{+}c_{k\sigma
}-\Delta _{N}\sum_{k\sigma }\sigma c_{k\sigma }^{+}c_{k+Q\sigma }\text{.}
\end{equation}%
Here, $\Delta _{k}^{N}=\Delta _{N}$ is isotropic, representing the strength
of the SDW. Its value can be in principle evaluated self-consistently with a reduced Coulomb repulsion $U$ in the mean-field level\cite{Luo-PRL05}. Here, we treat it as an independent parameter, determined experimentally. The chemical
potential $\mu $ is fixed by the particle conservation. The normal and
anomalous Green's functions are both $2\times 2$ matrices defined as $\hat{G}%
_{k\sigma }=-\langle T_{\tau }\psi _{k\sigma }(\tau )\psi _{k\sigma
}^{\dagger }\rangle $, and $\hat{F}_{k}=-\langle T_{\tau }\psi
_{-k\downarrow }(\tau )\psi _{k\uparrow }^{T}\rangle $, where $\psi
_{k\sigma }=(c_{k\sigma },c_{k+Q\sigma })^{T}$. The transversal spin
susceptibility under the random phase approximation, also a $2\times 2$
matrix, is expressed as $\hat{\chi}_{q}=\frac{\hat{\chi}_{q}^{0}}{1-U\hat{%
\chi}_{q}^{0}}$ with $U$ the above introduced reduced Coulomb repulsion. $%
\hat{\chi}_{q}^{0}=-\sum (\hat{G}_{k\downarrow }\hat{G}_{k+q\uparrow }+\hat{F%
}_{k}\hat{F}_{k+q}^{\ast })$ is the bare spin susceptibility, $k\equiv
(k,\omega )$.

In order to extract the role of superconductivity on the magnetic
excitations, we adopt the difference of spin susceptibility near $\mathbf{Q}$
between the SC and normal state as $I_{\mathbf{Q}}(\omega )=\int_{\Omega }%
\left[ \Im \chi _{q}^{SC}(\omega )-\Im \chi _{\mathbf{q}}^{NM}(\omega )%
\right] dq$, consisting with the experimental measurements\cite%
{Zhao-PRL07,Wilson-Nature06,Wilson-PNAS07}. The integration is restricted
within a small window of $0.01\pi \times 0.01\pi $ centered at $\mathbf{Q}$
point. The main features are insensitive to the selected integral region. In
numerics, the dressed TB parameters up to fourth nearest
neighbors, neglecting $t_{z}$, are adopted\cite{Markiewicz}. It should be
emphasized that our results are indeed not sensitive to the selected hoping
constant. In the electron-doped cuprates, we only focus on $x=0.15$, near
the optimal doping. The temperature is fixed at $3K$. To highlight the
resonance, a broaden factor $\Gamma =0.2meV$ is adopted.

\begin{figure}[btp]
\centering
\includegraphics[width=3.6in]{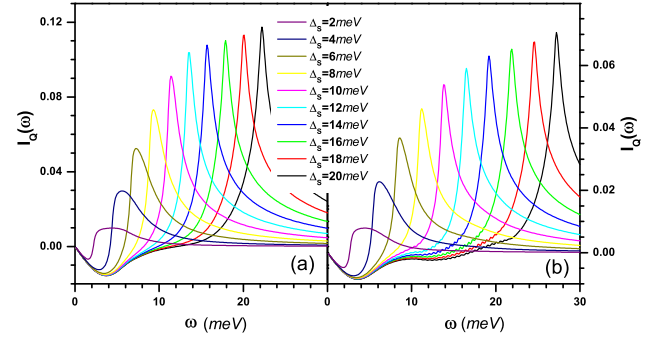}
\caption{(Color online) Low energy dependence of $I_{\mathbf{Q}}(\protect%
\omega)$ for different SC gap in hole-doped cuprates. (a) $x=0.10$, $%
\Delta_{N}=200meV$, $U=2.9t$; (b) $x=0.15$, $\Delta_{N}=150 meV$, $%
U=2.6t$. The dressed TB parameters are fitted from $LSCO$.
\cite{Markiewicz}}
\label{f.1}\vspace{-0.0in} \hspace{-0.0in}
\end{figure}

\begin{figure}[tbp]
\centering\includegraphics[width=3.6in]{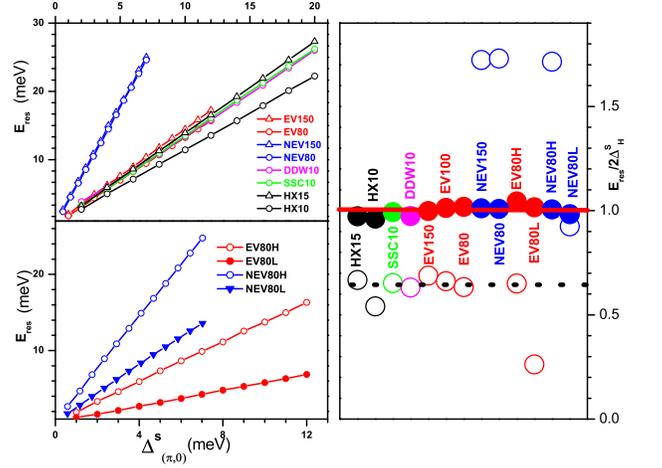}
\caption{(Color online) (a), and (b) Resonance energy $E_{res}$ as functions
of SC gap $\Delta^{S}_{(\protect\pi,0)}$ for monotonic and nonmonotonic
\emph{d}-wave pairing. (c) The suggested universal scaling $E_{res}/(2\Delta^{S}_{H})\sim1.0$ (Red solid line).
Hollow circles are for $E_{res}/(2\Delta^{S}_{(\pi,0)})$, and solid circles are the
renormalized data $E_{res}/(2\Delta^{S}_{H})$. For comparison, the experimental proposed universal
scaling $E_{res}/(2\Delta^{S}_{(\pi,0)})\sim0.64$ are also
shown (black dotted line). Other notations: 'HX15', and 'HX10' are for
hole-doped, $x=0.15$, and $x=0.1$. 'EV80', 'EV100', and 'EV150' are for
electron-doped with monotonic \emph{d}-wave, $\Delta_{N}=80meV$, $100meV$,
and $150meV$, respectively. 'NEV' is for nonmonotonic \emph{d}-wave, '+/-'
corresponds to the high/low-energy resonance as described in the text.
'DDW10' and 'SSC10' are data from DDW and SSC model with $x=0.1$.}
\label{f.3}
\end{figure}

\emph{Resonance and Fermi surface topology}
In Fig.~\ref{f.1}, two typical $I_{\mathbf{Q}}(\omega )$ in hole-doped
cuprates are shown. The spin gap emerges at low enough energy region, reflecting the fact that the
magnetic excitations are suppressed due to the opening of SC gap. $I_{%
\mathbf{Q}}(\omega )$ then increases quickly and reaches its maximum at
given energy, where the resonance $E_{res}$ is experimentally defined.
These low-energy features are qualitatively consistent with the INS
measurements. The resonance energy $E_{res}$ enhances with increasing SC gap $%
\Delta _{S}$, well consisting with experimental measurements. Our theoretical
data are even quantitatively comparable to the INS measurements\cite%
{Yu-NPHY09}. In optimal doping (Fig.~\ref{f.1}(b)), the estimated $%
E_{res}\sim 22meV$ for $\Delta _{S}=16meV$, well within the error bar in $%
LSCO$. In fact, the resonance energies exhibit a linear dependence on given
SC gap (Fig.~\ref{f.3}). In optimal doping case, this ratio of the linear
scaling $E_{res}/2\Delta^{AN} _{S}=0.66$, roughly consisting with the
measured data $0.51\pm 0.1$ and experimental proposed universal scaling $%
0.64 $\cite{Yu-NPHY09}. When the strength of SDW enhances, for example, $%
\Delta _{N}=200meV$ at $x=0.1$ (Fig.~\ref{f.1}(a)), the linear dependence
remains, but the ratio $E_{res}/2\Delta ^{AN}_{S}$ weakens
down to $0.54$.

\begin{figure}[tbp]
\centering\includegraphics[width=3.6in]{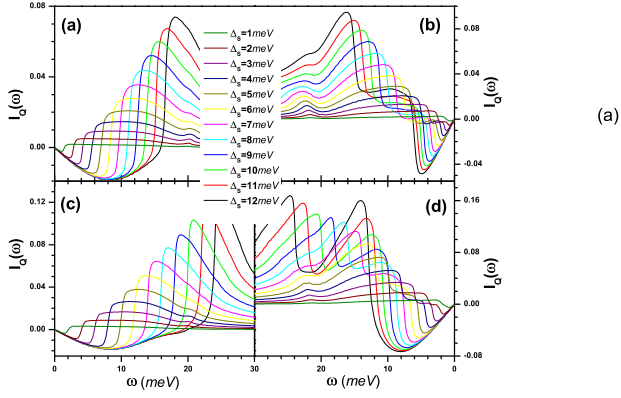}
\caption{(Color online) Low energy dependence of $I_{\mathbf{Q}}(\omega)$ for different SC gap in electron-doped cuprates. Upper panels are
for the monotonic \emph{d}-wave, and low panels are for the nonmonotonic
\emph{d}-wave. $\Delta_{N}=150meV$, $U=3.6t$ in
(a) and (c), $\Delta_{N}=80meV$, $U=3.2t$ in (b) and (d). The dressed
TB parameters are fitted from $NCCO$\protect\cite{Markiewicz}.}
\label{f.2} 
\end{figure}

The similar linear scaling can also be found in the electron-doped
cuprates both in the monotonic and nonmonotonic \emph{d}-wave cases as shown
in Fig.~\ref{f.2}. This suggests the linear dependence of $E_{res}$ on $%
\Delta _{S}$ is a universal features in the cuprates. The ratio
of $E_{res}/2\Delta^{AN} _{S}$ is about $0.69$ for $\Delta _{N}=150meV$
in case of the monotonic \emph{d}-wave, approximately approaching to the
suggested value $0.64$\cite{Yu-NPHY09}. In contrast, $E_{res}/2\Delta^{AN} _{S}$ is about $1.73$ in case of the nonmonotonic \emph{d}-wave,
exhibiting the strong deviations from $0.64$. We notice that the intensity
of $I_{\mathbf{Q}}(\omega )$ weakens with decreasing $\Delta _{S}$. This is
well consistent with INS measurements on $PLCCO$, where the external
magnetic field is applied to suppress the superconductivity\cite{Wilson-PNAS07}.

Interestingly, when the strength of SDW is reduced down to $80meV$, a weak
but visible low-energy peak emerges below $E_{res}$, especially for the
stronger superconductivity (Fig.~\ref{f.2}(b) and (d)). This is well
comparable with the recent INS data measured by Yu \emph{et al.} on the
optimal doped $NCCO$\cite{Yu-PRB10}, where a low-energy peak at $4.5meV$ is
discovered. However, the low-energy peak is absent in another INS
measurements on optimal doped $NCCO$\cite{Zhao-PRL07}. This may be due to
not well oxygen annealing in the latter sample, which enhancing the strength
of SDW. We also analyze the data with $\Delta _{N}=100meV$ (not shown),
where the low-energy resonance peak is almost invisible unless strong enough
SC gap is applied. Moreover, the low-energy resonance peak is also not found
in optimal doped $PLCCO$, where the SDW is much enhanced due to low doping
density ($x=0.12$). This is consistent with the case of $\Delta _{N}=150meV$.
We believe the present resonance with two peaks can be further discovered
in the slightly overdoped $n-$type cuprates. More INS measurements are
expected to check it.

As demonstrated above, our numerical data show that the distinct
ratio can be found in the respective case as shown in Fig.~\ref{f.3}(a).
Typically, $E_{res}/2\Delta^{AN}_{S}$ is about $0.5\sim 0.7$ in the
hole-doped cuprates, $0.6\sim 0.7$, and $1.7$ for high-energy resonance in the electron-doped cuprates with monotonic, and
nonmonotonic \emph{d}-wave, respectively. While for the low energy resonance, the ratio ranges from $0\sim 0.55$, and $0\sim
0.9$ in case of the monotonic, and nonmonotonic \emph{d}-wave. This ratio
can be even higher when the SDW is further suppressed. The experimentally
universal scaling $E_{res}/2\Delta^{AN}_{S}\sim 0.64$ seems to be
insufficient to cover the whole situations. Whether there still exists a universal scaling between the resonance energy and SC gap, and the intrinsic relation between the resonance and superconductivity are then naturally put forward.

\begin{figure}[tbp]
\centering\includegraphics[width=3.6in]{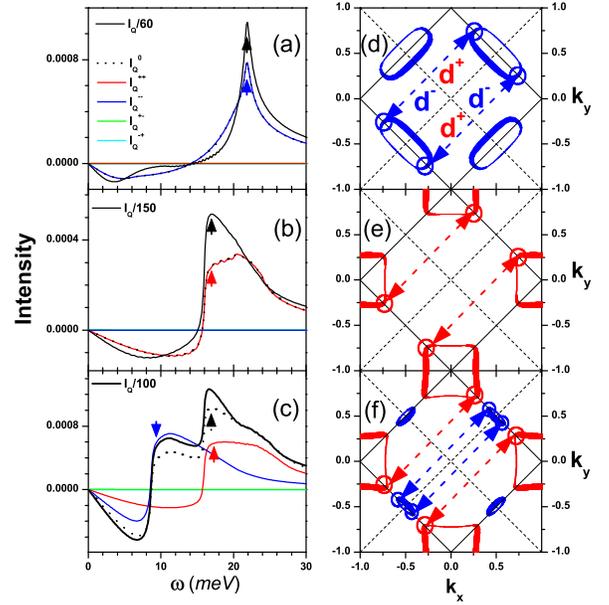}
\caption{(Color online) Left panels: $I_{\mathbf{Q}}(\protect\omega)$ and $%
I_{\mathbf{Q}}^{0}$ at given SC gap $\Delta_{S}$, together with the four
components as described in the text. (a) hole-doped case with $x=0.15$, $%
\Delta_{N}=150meV$, and $\Delta_{S}=16meV$. (b), and (c) electron-doped case
with nonmonotonic \emph{d}-wave, $\Delta_{N}=150meV$, and $80meV$,
respectively. $x=0.15$, $\Delta_{S}=8meV$. Right panels are the
corresponding Fermi surface. The thickness denotes the intensity. Hot spots
are denoted by the hollow circles, and the arrows represent the $\mathbf{Q}$%
-scattering. The thin solid lines are the magnetic Brillouin zone, and the
thin dotted lines divide the full Brillouin zone into positive and negative
region for \emph{d}-wave superconductivity as denoted in (d).}
\label{f.4}
\end{figure}

To understand these issues, we have to analyze the magnetic response further.
IN fact, the main features of the bare spin susceptibility difference $I_{\mathbf{Q}%
}^{0}(\omega )$ are almost the same as the RPA one as shown in the left
panels in Fig.~\ref{f.4}, but with much reduced intensity. For simplicity,
we directly analyze $I_{\mathbf{Q}}^{0}(\omega )$ instead. It includes four
components: the intra-band components $I_{\mathbf{Q}}^{++}$ and $I_{\mathbf{Q%
}}^{--}$, the inter-band components $I_{\mathbf{Q}}^{+-}$ and $I_{\mathbf{Q}%
}^{-+}$. In hole-doped cuprates, $I_{\mathbf{Q}}^{0}(\omega )$ fully comes from the
intra-band components $I_{\mathbf{Q}}^{--}$ (Fig~\ref{f.4}(a)). As well
known that the Fermi surface is a hole-pocket due to the lower band ($-$)
crossing the Fermi energy in presence of SDW. It loses most intensity beyond
the magnetic Brillouin zone (Fig~\ref{f.4}(d)), resulting the well known
'arc'-type Fermi surface. When we focus on the difference of integration
nearby $\mathbf{Q}=(\pi ,\pi )$, the scattering between the two hot spots
contributes the most intensity as denoted by arrows. On the other hand, the
SC order parameter changes sign between the two hot spots, which is the
essence of the resonance as stressed previously\cite{Bulut}. In electron-doped
cuprates, only electron-pocket near antinodes can be found for strong SDW
(Fig.~\ref{f.4}(e)), which is induced by the upper bands ($+$). $I_{\mathbf{Q%
}}^{0}(\omega ) $ fully comes from the intra-band component $I_{\mathbf{Q}%
}^{++}$ due to the dominate scattering between the two hot spots. Therefore,
only a single resonance peak can be found in the above two cases. However,
they come from respective component.

The situation changes when SDW are weakened in the electron-doped cuprates.
The hole and electron pocket emerge simultaneously (Fig.\ref{f.4}(f)),
consisting with the large three-piece Fermi surface structure found in
optimal doped $NCCO$ \cite{Armitage-PRL01}. Now, there exists two types of
scattering between the hot spots. Both intra-band components contribute to $%
I_{\mathbf{Q}}^{0}(\omega )$ as shown in Fig.~\ref{f.4}(c). The low-, and
high-energy resonance comes from $I_{\mathbf{Q}}^{--}$, and $I_{\mathbf{Q}%
}^{++}$, respectively. Therefore, the two resonances are related to the $%
B_{1g}/B_{2g}$ found in the Raman scattering\cite{Qazilbash-PRB05}. The
low-energy resonance can be covered by the high-energy resonance when SDW
enhances slightly, for example $\Delta _{N}=100meV$ as we mentioned above,
where the component $I_{\mathbf{Q}}^{--}$ is too small to be discovered.
This is a possible reason for the absence of the low-energy resonance even
in the optimal doped $NCCO$ as discovered by Zhao \emph{et al.}\cite%
{Zhao-PRL07}. The two resonances may also be coincident when SDW is weak
enough due to proximity of the two types of hot spots, which is expected to
discovered in the orverdoped \textit{n}-type cuprates. Therefore, the
features of resonance can be well understood by the scattering between
the hot spots induced by the Fermi surface topology.

Reminding the scattering between the hot spots contributes mostly as
mentioned in the above analysis. We therefore propose to adopt $%
E_{res}/2\Delta ^{H}_{S}$ instead, where $\Delta ^{H}_{S}$ is the magnitude of the SC gap at
the hot spot. There are two types of hot spot in the SDW suppressed
electron-doped cuprates. After renormalization, a new scaling with $%
E_{res}/2\Delta ^{H}_{S}\sim 1.0$ is well established. This universal
scaling is independent on the details, whatever the monotonic/nonmonotonic
\emph{d}-wave, hole-/electron-doping, high/low energy resonance, and the
selected dressed TB parameters. In fact, it is also model
independent if the Fermi surface topology is established. In Fig. \ref{f.3},
we also present the results obtained from DDW and SSC introduced before. The
same ratio further manifests that the resonance is rather a consequence of
Fermi surface topology. Further INS
measurements on various cuprates are expected to justify this universal
scaling.

\emph{Resonance and pairing symmetry} To understanding the intrinsic
relation between the resonance and superconductivity, we consider two other
singlet pairing symmetries: the on-site \emph{s}-wave $\Delta
^{k}_{S}=\Delta $, and extended \emph{s}-wave $\Delta ^{k}_{S}=\frac{1}{2}%
\Delta (\cos k_{x}+\cos k_{y})$. The former is typically found in the
conventional superconductor, while the latter is proposed in the recent
discovered Fe-pnictides, producing the so-called $s^{\pm }$-wave nature\cite%
{Stewart}. No resonance can be found in both two types of pairing
symmetry due to the lack of the sign changing of the SC gap between the two hot spots. Therefore, the resonance in cuprates is closely linked to the \emph{d}-wave pairing symmetry.

The monotonic \emph{d}-wave pairing is a consensus in the hole-doped
cuprates. However, it is controversial in the electron-doped cuprates. Both
the monotonic and nonmonotonic \emph{d}-wave are proposed to explain the ARPES data\cite{Armitage-RMP}. The resonance
in $x=0.12$ doped $PLCCO$ is about $11$meV\cite{Wilson-Nature06}. This means
that the SC gap at the antinode should be $7.5meV$, and $3meV$ in the
monotonic, and nonmonotonic \emph{d}-wave pairing providing $\Delta
_{N}=150meV$, respectively. Matsui \emph{et al.} revealed that $\Delta
^{AN}_{S}\sim 2.0meV$ at $8K$ by the ARPES measurements\cite{Matsui}.
Therefore, the pairing in electron-doped cuprates is more likely a
nonmonotonic \emph{d}-wave. Furthermore, the experimental low-energy
resonance peak in optimal doped $NCCO$ is about $E_{res}=4.5meV$ \cite%
{Yu-PRB10}. It roughly corresponds to the antinodal SC gap about $7\sim 8mev$%
, and $2.2meV$ in the monotonic, and nonmonotonic \emph{d}-wave in case $%
\Delta _{N}=80meV$. Compared with the Raman scattering measurements\cite%
{Blumberg}, the gap along the antinodal direction is merely about $3meV$,
corresponding to $\Delta ^{AN}_{S}\sim 2.2meV$. We again confirm that
the SC pairing in electron-doped cuprates is the nonmonotonic \emph{d}-wave.
The corresponding high-energy peak is about $8.5meV$ for $\Delta _{N}=80meV$%
, consisting with the data obtained by Zhao \emph{et al}.\cite{Zhao-PRL07}.
In fact, the high energy resonance changes little when SDW enhances. In contrast, the high-energy resonance is merely $6.5meV$ obtained by Yu \emph{et al.}, which may underestimate the high energy component in their analysis. More INS measurements are expected to clarify the
discrepancy. Therefore, the INS technique can be used to verify the pairing
symmetry, and even the details.

The similar scaling $E_{res}/2\Delta
\sim 0.64$ had also been suggested in the iron-based superconductor by INS\cite%
{Zhao-NPHY11}. More recently, a collective boson mode with $E_{res}\sim
2\Delta $ is found by the scanning tunnelling spectroscopy\cite{HHW-NP13}.
The hole-pocket, and electron-pocket emerge near $(0,0)$, and $(\pi ,\pi )$
point, where the $s^{+}$ and $s^{-}$-wave locate, respectively\cite{Stewart}%
. Therefore, the dominate contributions are expected to come from the
inter-band components. Furthermore, the iron-based superconductors are thought to
be analog to the cuprates after a gauge mapping\cite{Hu-PRX}. Our
analysis on the cuprates may be further applied to the iron-based superconductors.

In summary, the resonance and its intrinsic links to the
superconductivity are studied in cuprates. The main features discovered
experimentally are well established. The resonance energy exhibits linear
dependence on the SC gap. A low-energy resonance develops in the
electron-doped cuprates when SDW is suppressed. The respective resonance
comes from the different origins in hole-/electron-doped cuprates, and the
low-/high-energy resonance in the electron-doped cuprates. We further propose a universal scaling $E_{res}/2\Delta
^{H}_{S}\sim 1.0$ instead of the experimentally suggested $E_{res}/2\Delta^{AN}_{S}\sim 0.64$, irrespective of the detailed selections. Present
work strongly suggested that the resonance is rather a consequence of \emph{d%
}-wave pairing nature and Fermi surface topology than the glue of the
superconductivity. Based on our
analysis, the pairing symmetry in electron-doped cuprates is more likely the
nonmonotonic \emph{d}-wave pairing. Therefore, the resonance in INS
measurements is not only related to the pairing symmetry, but also to the
detailed form.

This work was supported by NSFC Projects No. 11274276, and A Project Funded
by the Priority Academic Program Development of Jiangsu Higher Education
Institutions. CD Gong acknowledges 973 Projects No. 2011CBA00102. HQ Lin
acknowledges MOST 2011CB922200 and NSFC U1230202.



\begin{thebibliography}{99}
\bibitem{Armitage-RMP} N. P. Armitage, P. Fournier, and R. L. Greene, Rev.
Mod. Phys. \textbf{82}, 2421 (2010).

\bibitem{Jin-Nature11} K. Jin \emph{et al.}, Nature \textbf{476}, 73 (2011).

\bibitem{FujitaRev} M. Fujita \emph{et al.}, J. Phys. Soc. Jpn. \textbf{81},
011007 (2012).

\bibitem{Zhao-NPHY11} J. Zhao \emph{et al.}, Nature Phys. \textbf{7}, 719
(2011).

\bibitem{Fujita-PRL08} M. Fujita \emph{et al.}, Phys. Rev. Lett. \textbf{101}%
, 107003 (2008).

\bibitem{Yamada-PRL03} K. Yamada \emph{et al.}, Phys. Rev. Lett. \textbf{90}%
, 137004 (2003).

\bibitem{Zhao-PRL07} J. Zhao \emph{et al.}, Phys. Rev. Lett. \textbf{99},
017001 (2007).

\bibitem{Wilson-Nature06} S. D. Wilson \emph{et al.}, Nature \textbf{442},
59 (2006).

\bibitem{Wilson-PNAS07} S. D. Wilson \emph{et al.}, PNAS \textbf{104}, 15259
(2007).

\bibitem{Yu-NPHY09} G. Yu \emph{et al.}, Nature Phys. \textbf{5}, 873
(2009), and refered therein.

\bibitem{Yu-PRB10} G. Yu \emph{et al.}, Phys. Rev. B \textbf{82}, 172505
(2010).

\bibitem{Niestemski-Nature07} F. C. Niestemski \emph{et al.}, Nature, \textbf{450}, 1058 (2007).

\bibitem{Feng} L. Cheng, and S. P. Feng, Phys. Rev. B \textbf{77}, 054518
(2008).

\bibitem{Ismer-PRL07} J.-P. Ismer \emph{et al.}, Phys. Rev. Lett, \textbf{99}%
, 047005 (2007).

\bibitem{LJX-PRB03} J. X. Li, J. Zhang, and J. Luo, Phys. Rev. B \textbf{68}%
, 224503 (2003).

\bibitem{Kruger-PRB07} F. Kr\"{u}ger \emph{et al.}, Phys. Rev. B \textbf{76}%
, 094506 (2007).

\bibitem{HY} H. Y. Zhang \emph{et al.}, J. Phys. Condens. Matter \textbf{25}%
, 155603 (2013).

\bibitem{Norman-PRB07} M. R. Norman \emph{et al.}, Phys. Rev. B \textbf{76},
174501 (2007).

\bibitem{Das-PRB85} T. Das, R. S. Markiewicz, and A. Bansil, Phys. Rev. B
\textbf{85}, 064510 (2012).

\bibitem{Greco-PRL09} A. Greco, Phys. Rev. Lett. \textbf{103}, 217001
(2009); M. Bejas \emph{et al.}, Phys. Rev. B \textbf{82}, 014514 (2011).

\bibitem{Chakravarty-PRB01} S. Chakravarty \emph{et al.}, Phys. Rev. B
\textbf{63}, 094503 (2001).

\bibitem{ZLG-JPSJ13} Y. Zhou, H.\ Q. Lin, and C. D. Gong, J. Phys. Soc. Jpn.
\textbf{82}, 034702 (2013).



\bibitem{Matsui} H. Matsui \emph{et al.}, Phys. Rev. Lett., \textbf{95},
017003 (2005).

\bibitem{Blumberg} G. Blumberg \emph{et al.}, Phys. Rev. Lett., \textbf{88},
107002 (2002).

\bibitem{Luo-PRL05} H. G. Luo and T. Xiang, Phys. Rev. Lett. \textbf{94},
027001 (2005).

\bibitem{Markiewicz} R. S. Markiewicz \emph{et al.}, Phys. Rev. B \textbf{72}%
, 054519 (2005).

\bibitem{Bulut} N. Bulut and D. J. Scalapino, Phys. Rev. B \textbf{53}, 5149
(1996).

\bibitem{Armitage-PRL01} N. P. Armitage \emph{et al.}, Phys. Rev. Lett.
\textbf{87}, 147003 (2001).

\bibitem{Stewart} G. R. Stewart, Rev. Mod. Phys. \textbf{83}, 1589 (2011).

\bibitem{Qazilbash-PRB05} M. M. Qazilbash \emph{et al.}, Phys. Rev. B
\textbf{72}, 214510 (2005).

\bibitem{HHW-NP13} Z. Y. Wang \emph{et al.}, Nature Phys. \textbf{9}, 42
(2013).

\bibitem{Hu-PRX} J. P. Hu, and N. N. Hao, Phys. Rev. X \textbf{2}, 021009
(2012).






\end{thebibliography}
\end{document}